\documentclass[pdflatex,sn-nature]{sn-jnl}

\usepackage{graphicx}%
\usepackage{multirow}%
\usepackage{amsmath,amssymb,amsfonts}%
\usepackage{amsthm}%
\usepackage{mathrsfs}%
\usepackage[title]{appendix}%
\usepackage{xcolor}%
\usepackage{textcomp}%
\usepackage{manyfoot}%
\usepackage{booktabs}%
\usepackage{algorithm}%
\usepackage{algorithmicx}%
\usepackage{algpseudocode}%
\usepackage{listings}%
\usepackage[numbers]{natbib}
\usepackage{multicol}
\usepackage{microtype}

\definecolor{red}{RGB}{255,0,0} 

\usepackage[final]{changes} 
\usepackage{changes}\theoremstyle{thmstyletwo}%

\begin{document}
\title[Article Title]{Room temperature Purcell enhanced single erbium ions in silicon-carbide-on-insulator microring resonators}

\author[1]{\fnm{Joshua} \sur{Bader}}
\author[2]{\fnm{Shin-ichiro} \sur{Sato}} 
\author[3]{\fnm{Jeffrey C.} \sur{McCallum}} 
\author[4]{\fnm{Ruixuan} \sur{Wang}}
\author[3, 5]{\fnm{Shao Qi} \sur{Lim}} 
\author[6]{\fnm{Alexey} \sur{Lyasota}}
\author[5]{\fnm{David} \sur{Broadway}}
\author[5]{\fnm{Brett C.} \sur{Johnson}}
\author[6]{\fnm{Sven} \sur{Rogge}}
\author[4]{\fnm{Qing} \sur{Li}} 
\author[1]{\fnm{Stefania} \sur{Castelletto}} 


\affil[1]{\orgdiv{School of Engineering}, \orgname{RMIT University}, \orgaddress{\city{Melbourne}, \postcode{3000}, \state{VIC}, \country{Australia}}}
\affil[2]{\orgdiv{National Institutes for Quantum Science and Technology}, \orgaddress{\city{Takasaki, Gunma}, \postcode{370-1292}, \country{Japan}}}
\affil[3]{\orgdiv{Centre for Quantum Computation and Communication Technology, School of Physics}, \orgname{The University of Melbourne}, \orgaddress{\city{Melbourne}, \postcode{3010}, \state{VIC}, \country{Australia}}}
\affil[4]{\orgdiv{Electrical and Computer Engineering}, \orgname{Carnegie Mellon University}, \orgaddress{\city{Pittsburgh}, \postcode{15213}, \state{PA}, \country{USA}}}
\affil[5]{\orgdiv{School of Science}, \orgname{RMIT University}, \orgaddress{\city{Melbourne}, \postcode{3001}, \state{VIC}, \country{Australia}}}
\affil[6]{\orgdiv{Centre of Excellence for Quantum Computation and Communication
Technology, School of Physics}, \orgname{University of New South Wales}, \orgaddress{\city{Sydney}, \postcode{2052}, \state{NSW}, \country{Australia}}}


\abstract{\added{Spin-carrying single-photon emitters} \deleted{residing} \added{operating in} the telecommunication C-band (1530-1565nm) are prime candidates for integrated spin-photon interfaces, offering seamless compatibility with existing fiber-optic infrastructure, an essential component for future quantum \deleted{information processing} \added{networks}. 
\added{In this context}, erbium-dopants ($\text{Er}^{3+}$) \added{are particularly compelling due to their exceptional emitter properties, including small spectral diffusion and long spin coherence times. However, their low C-band photon-emission rate and operation at cryogenic temperatures has limited the realization of this technology.} 

In this work, we demonstrate fully integrated single-photon \added{emission from} an ion implanted $\text{Er}^{3+}$-\added{defect} embedded into a 4H-silicon-carbide-on-insulator \added{(4H-SiCOI)} microring resonator operating at \added{room temperature}. \deleted{We observe a mode overlap between the microring and $\text{Er}^{3+}$-defect which led to the observation of} \added{By optimizing the mode overlap between the resonator and the $\text{Er}^{3+}$-defect, we achieved a } $\sim$70$\times$ Purcell enhancement and \added{recorded} small spectral diffusion of $\sim$54 MHz.  \added{We} further \added{characterize the $\text{Er}^{3+}$ single photon emission via photon correlation} \deleted{insights presented into observed} g$^{(2)}$-histograms \added{and investigate its performance under varying magnetic-field, demonstrating Zeeman splitting on single emitters}. }




\maketitle

\section*{Introduction}\label{sec1}
\begin{multicols}{2}

\added{Coupling single quantum emitters possessing an intrinsic spin degree of freedom to optical cavities enables  efficient spin-photon interfaces that bridge stationary qubits (electron or nuclear spins) and flying qubits (photons). Such interfaces constitute the fundamental building blocks of long-distance quantum networks~\cite{kimble2008quantum}, allowing quantum information stored in long-lived spins to be transferred over fiber-optic links. A wide range of platforms are being explored, including color centres in diamond~\cite{nguyen2019quantum,fischer2025spin}, silicon carbide (SiC)~\cite{christle2017isolated,babin2022fabrication,anderson2022five}, hexagonal boron nitride~\cite{stern2024quantum}, the T-centre in silicon \cite{higginbottom2022optical}, InGaAs quantum dots \cite{javadi2018spin}, donor spins in silicon (Si) \cite{mi2018coherent} and rare earth defects such as Erbium (Er$^{3+}$) in crystals \cite{gupta2025dual} and molecules~\cite{Weisshighresolution2025}. Despite their diverse physical advantages, the viability of these systems for large-scale quantum networks ultimately depends on their integration into complementary metal–oxide–semiconductor (CMOS) compatible photonic integrated circuits (PICs) that can be fabricated at wafer scale~\cite{dobinson2025electrically}.}

\added{Among these platforms,} optically addressable \added{spin qubits in 4H-SiC} combine long spin coherence times~\cite{anderson2022five} (typically in the ms-regime~\cite{widmann2015coherent,christle2015isolated}), single and indistinguishable photon emission \cite{morioka2020spin}, \added{telecom O-band compatibility \cite{cilibrizzi2023ultra} and electrical control of the spin-photon interface~\cite{niethammer2019coherent, anderson2019electrical,steidl2025single}. \deleted{as well as,}} 
Combined with the relative ease of defect-integration into SiC-on-insulator (SiCOI) and low-loss photonic integrated circuits (PICs)~\cite{lukin20204h}, in composition with the straightforward generation of these defects, which can be formed via electron irradiation~\cite{nagy2019high}, ion implantation~\cite{cilibrizzi2023ultra} and direct laser writing~\cite{chen2019laser}, positioning this spin system as a ideal platform for future quantum technology applications.

Especially spin–photon interfaces operating directly in the telecommunication bands avoid the need for frequency conversion in long-distance quantum communication. In this context, $\text{Er}^{3+}$ is particularly attractive: its ${}^\text{4}\text{I}_{\text{13/2}}$ $\rightarrow$ ${}^\text{4}\text{I}_{\text{15/2}}$ transition lies in the telecom C-band ($\sim$1540 nm) and is shielded by the 4f electronic shell, resulting in weak coupling to the host lattice~\cite{babunts2000properties} and narrow optical transitions. In particular, recent progress in Er$^{3+}$-doped silicon-on-insulator (SOI) PICs has shown Purcell enhanced single-photon emission~\cite{gritsch2022narrow, gritsch2023purcell}, optical spin manipulation with single-shot readout~\cite{gritsch2025optical} and compatibility with quantum memories~\cite{ramirez2024integrated}, constrained by the intrinsic two-photon absorption and smaller bandgap, issues that could be mitigated in 4H-SiCOI.

In this context, recent studies of $\text{Er}^{3+}$ in 4H-SiCOI revealed sub-megahertz homogeneous linewidths in weak ensembles~\cite{lyasota2025narrow}, attributed to reduced thermal quenching, stable lattice incorporation and low local strain. These measurements are typically conducted in a photoluminescence excitation (PLE) configuration where the excited-state manifold (${}^\text{4}\text{I}_{\text{13/2}}$) splits into seven crystal-field levels, while the ground-state manifold (${}^\text{4}\text{I}_{\text{15/2}}$) splits into eight, with a branching ratio of approximately 0.5, consistent with the formation of two dominant Er sites at cryogenic temperatures. At room temperature (RT), PLE-spectroscopy allows access to all available radiative transitions which are limited to the lowest lying ${}^\text{4}\text{I}_{\text{15/2}}$-ground state crystal field at cryogenic temperatures, potentially preserving spin coherence limited only by the emitter lifetime~\cite{ranvcic2018coherence}. However, the intrinsically long millisecond optical lifetime and dipole orientation along the in-plane crystallographic c-axis~\cite{bader2025photo} result in weak photon emission, which has hindered the observation of single-photon emission so far. This can be overcome by engineering the local density of optical states using high-quality-factor photonic cavities~\cite{joannopoulos2008molding,lukin20204h, cai2022octave} to achieve significant Purcell enhancement~\cite{purcell1995spontaneous} and efficient coupling to integrated waveguides.

Here, we demonstrate the integration of single $\text{Er}^{3+}$ defects in 4H-SiCOI coupled to high-Q microring resonators. The resonator is evanescently coupled to feed waveguides with grating couplers for on-chip excitation and photon collection into single-mode fibres. Utilizing resonant, modulated excitation, we observe single-photon emission  with an identified $\text{g}^{(2)}(\tau = 0)$ = 0.04 without background correction, confirming high-purity single-photon emission from individual $\text{Er}^{3+}$ defects and measure a spectral diffusion of 54.76 ± 1.8 MHz at RT. We achieve a maximum lifetime reduction factor of 114, corresponding to a Purcell factor of 69.8, exceeding previously reported values for divacancies ($\text{V}_{\mathrm{Si}}\text{V}_{\mathrm{C}}$)~\cite{crook2020purcell} in SiC and comparable to silicon-vacancies ($\text{V}_{\text{Si}}^{-}$) and $\text{Er}^{3+}$ photonic crystal cavity systems~\cite{bracher2017selective, gritsch2023purcell}. 
We further observe magnetic-field dependent behavior and coherent optical response, establishing $\text{Er}^{3+}$ in 4H-SiCOI as a scalable telecom-band spin-photon interface compatible with CMOS photonics.

\end{multicols}
\section*{Results}\label{sec2}
\subsection*{Photonic device characterization}\label{sec2}
\begin{multicols}{2}

\begin{figure}[H]
    \includegraphics[keepaspectratio, width=\textwidth]{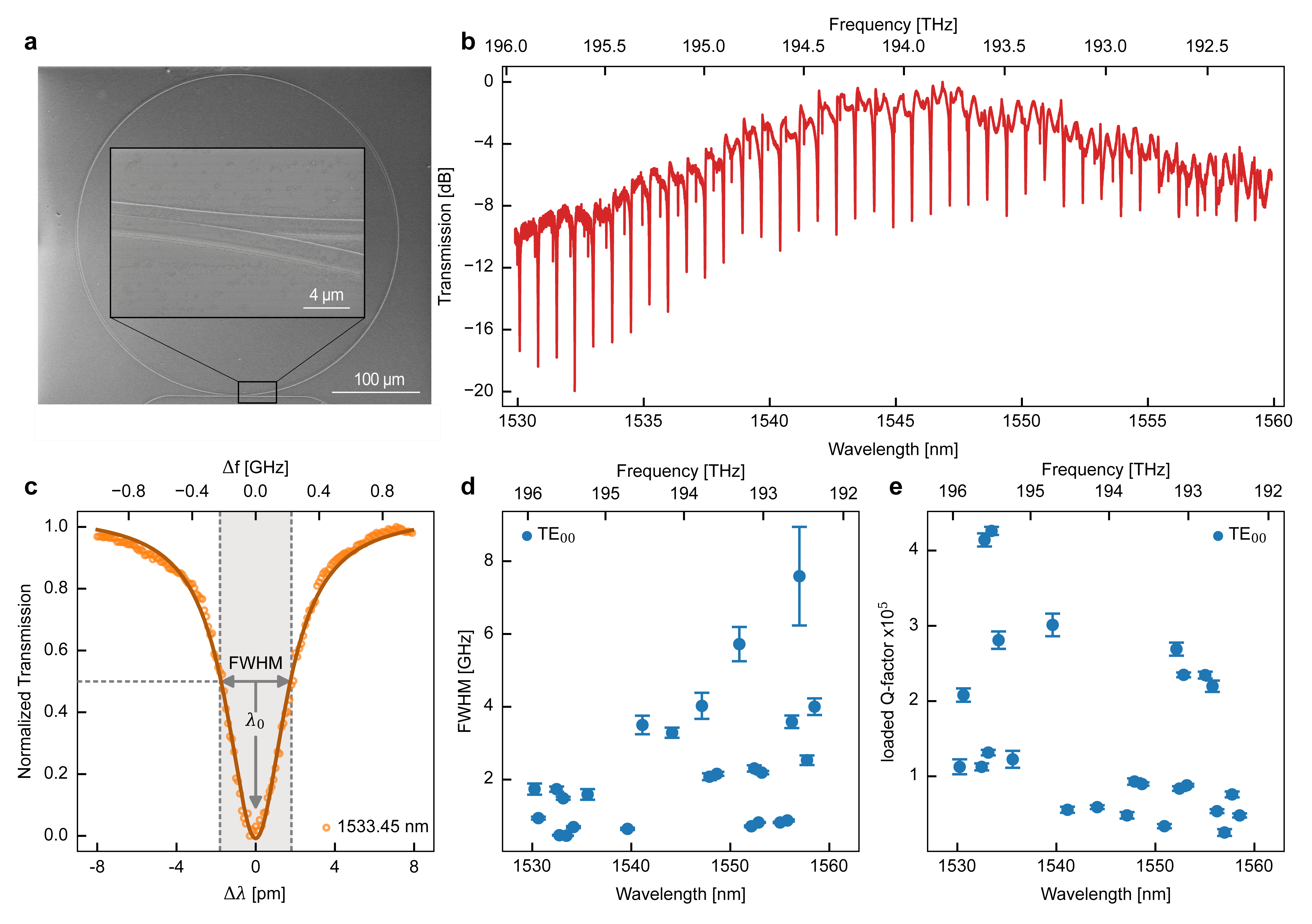}
    \caption{Characterization of the Er$^{3+}$-implanted 4H–SiCOI microring resonator: \textbf{a} SEM image of a representative microring resonator (scale bar 100 $\mu$m). The inset shows a 45\textdegree-tilted view of the microring–waveguide coupling region (scale bar 4 $\mu$m); \textbf{b} Spectral transmittance of the implanted and annealed device, showing sharp cavity resonances across the 1530-1560 nm wavelength range; \textbf{c} Narrowest observed $\text{TE}_{00}$ resonance exhibiting an full width half maximum (FWHM) of 458.85 ± 5.7 MHz at the identified center wavelength $\lambda_0$ where measured data (scattered) is fitted with a Lorentzian distribution (solid line);
    \textbf{d} Overview of the identified FWHMs over exhibited wavelengths from all $\text{TE}_{00}$ resonances shown in \textbf{b}; \textbf{e} Loaded quality factors (Q) extracted from all $\text{TE}_{00}$ resonances shown in \textbf{b}. Error bars represent fitting uncertainties. FWHMs were determined from Lorentzian fits. } 
    \label{fig1} 
\end{figure}

We begin with investigating the spectral response from the implanted ($1.2\times10^{16}$ $\text{Er}/\text{cm}^3$) and thermally annealed microring resonator (diameter $\sim$300 $\mu$m) where clear sidewalls and features are preserved after defect generation (see Methods), as shown in Fig. \ref{fig1}\textbf{a}. We identify clear sharp resonances distanced by the free spectral range \added{(FSR)} of $\sim$0.7 nm, as shown in Fig. \ref{fig1}\textbf{b}. Sharpest resonances exhibit full width half maxmimum (FWHM) values of 458.85 ± 5.7 MHz observable at center wavelength $\lambda_0$ = 1533.45 nm, as illustrated in Fig. \ref{fig1}\textbf{c} which point to a highest loaded Q-Factor of $\sim4.26\times 10^{5}$. 
Overviews of all identified FWHM and Q-Factors are presented in Fig. \ref{fig1}\textbf{d}, \textbf{e} (see SI I for each individual fit).
\added{Additionally,} the Er ion implantation fluence was swept between $5\times10^{10}$, $1\times10^{12}$ to $1\times10^{14}$ $\text{Er}/\text{cm}^2$ to study its impact on the spectral response after waveguide-fabrication (see Methods for ion implantation details). 
With this, we observed a deleterious spectral response with higher Er$^{3+}$-fluences (see SI II) in agreement with previous studies \cite{xu2021ultrashallow, liu2022photonic}.

\end{multicols}
\begin{multicols}{2}


\subsection*{Optical properties of Er-defects within 4H-SiCOI}\label{sec3}
\begin{figure}[H]
    \includegraphics[keepaspectratio, width=\textwidth]{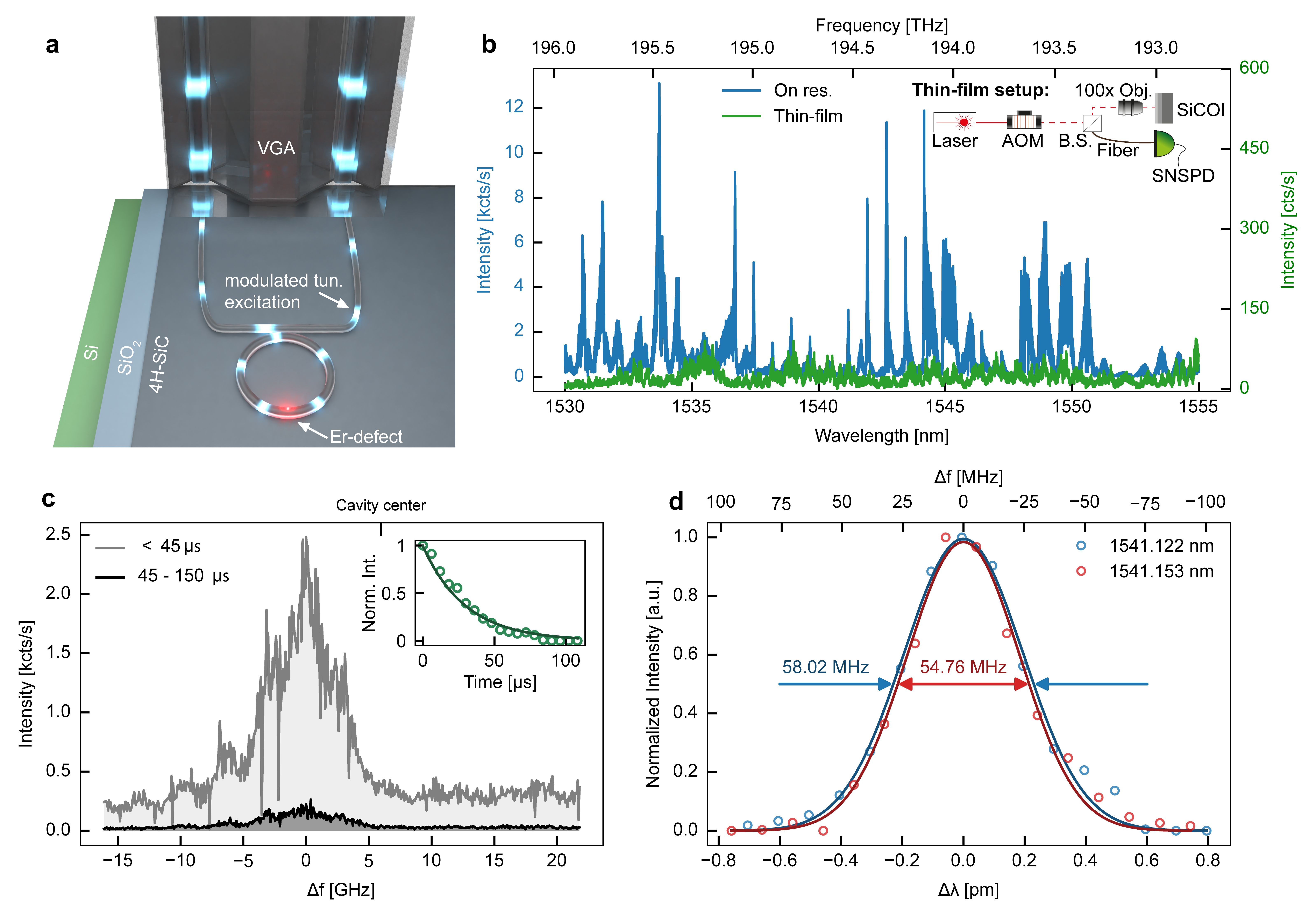}
    \caption{Optical properties from single Er$^{3+}$-defects within 4H-SiCOI: \textbf{a} Schematic measurement setup where modulated tunable excitation is coupled into the photonic device via a V-grooved fiber-array (VGA) and subsequently outcoupled and directed towards an SNSPD in a photoluminescence excitation configuration; \textbf{b} obtained PLE-spectrum from the $1.2\times10^{16}$ $\text{Er}/\text{cm}^{3}$ implanted ring resonator conducted via the grating coupler (blue) and in a classical confocal configuration (green), where the inset illustrates the utilized free-space PLE setup for `thin-film' characterization indicating an excitation-application via a 10:90 beam splitter (B.S.); \textbf{c} Lifetime-dependent PLE-section centered at 1541.172 nm detected within the first 45 $\mu$s after excitation-extinction (grey) and between 45-150 $\mu$s (black). The label on the top axis illustrates the observed cavity center exhibited by the microring while the inset represents an obtained lifetime transient from a peak situated at approx. -6.94 GHz detuning; \textbf{d} spectral diffusion overlap for selected observed peaks with narrowest linewidth identified at 54.76 $\pm$ 1.8 MHz. }\label{fig2} 
\end{figure}

Next, we utilize this microring to address the optically active Er$^{3+}$-dopants via modulated tunable excitation \deleted{embedded} \added{guided} into the PIC via the grating couplers\deleted{exclusively}, as shown schematically in Fig. \ref{fig2}\textbf{a}. With this technique, we excite the Er$^{3+}$-dopants perpendicular to the c-axis of the crystal, in parallel to the direction of the Er$^{3+}$ emission dipole\deleted{from the defect}~\cite{bader2025photo}.  \deleted{which further allows us} \added{The laser excitation and waveguide propagation modes effectively} guide the Er$^{3+}$ \added{single photon} emission via the fabricated waveguides and efficiently outcouple them \deleted{this} \deleted{in a efficient way} towards a \added{Superconducting Nanowire Single Photon Detector} (SNSPD) \deleted{for detection purposes} (see Methods). This enabled us to perform a PLE scan within the microring \added{resonator} \deleted{ranging} \added{spanning} from 1530 nm to 1555 nm, where most of the known Er$^{3+}$ transition lines reside \cite{lyasota2025narrow}, referenced as `on resonance' in Fig. \ref{fig2}\textbf{b}. 

Furthermore, similar investigations were performed in unpatterned thin-film 4H-SiCOI (here referred as `thin-film') which are obtained by \added{a high numerical aperture objective focusing the laser excitation vertically on the material} (see inset of Fig. \ref{fig2}\textbf{b}). 
By overlapping both obtained scans, we identified several enhanced Er$^{3+}$ transitions as well as an overlap of the cavity resonances provided by the microring resonator.
By comparing individual arrival-time intervals for a specific scanned section centered at 1541.172 nm from Fig. \ref{fig2}\textbf{b}, we can identify a dominant inhomogeneous peak with multiple arising Er$^{3+}$-transition \deleted{sharp} lines within the first 45 $\mu$s after excitation-extinction (as shown in Fig. \ref{fig2}\textbf{c}), which decay explicitly fast, indicating successful coupling to the resonator propagating mode~\cite{ulanowski2022spectral, gritsch2023purcell}.
A small background signal was also identified when time-intervals of 45 $\mu$s to 150 $\mu$s are considered, which could arise from defects \added{activated} within the feed-waveguide since the sample was uniformly implanted with Er$^{3+}$-dopants. 

In addition, small spectrally diffused Er$^{3+}$-transition lines were \added{measured} at 58.02 $\pm$ 1.8 MHz and 54.76 $\pm$ 1.8 MHz observed at 1541.122 nm and 1541.153 nm at RT respectively, with an overlap shown in Fig. \ref{fig2}\textbf{d}. These values \deleted{remain} \added{are} smaller than previous findings \deleted{with determined values} of 79 MHz \added{for Er$^{3+}$ transition at $\sim$1535.99 nm} in \added{SOI}~\cite{gritsch2023purcell} observed at cryogenic temperature $<$4K. 

\end{multicols}
\begin{multicols}{2}

\subsection*{Purcell enhancement}\label{sec4}
\begin{figure}[H]
    \centering
    \includegraphics[keepaspectratio, width=0.48\textwidth]{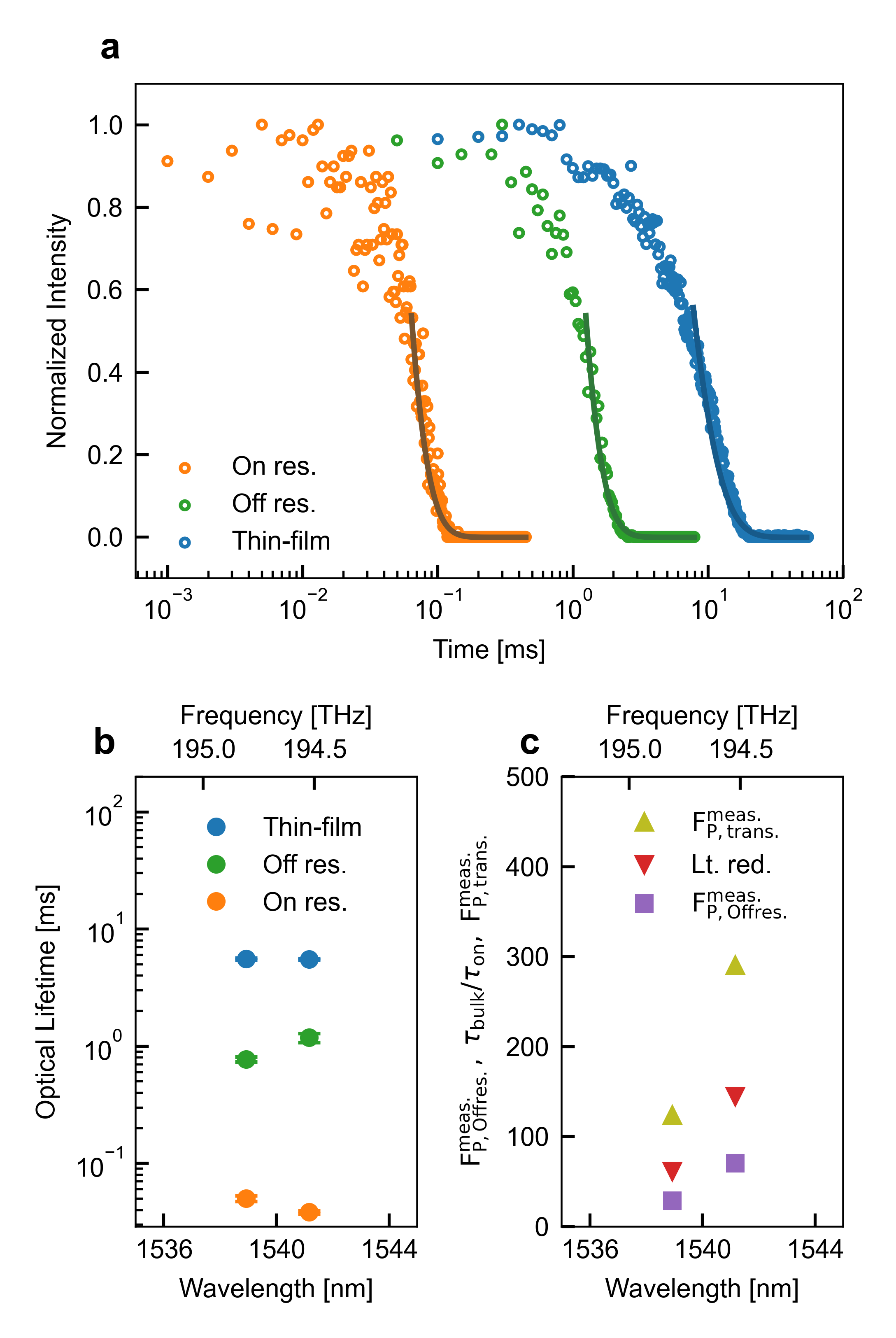}
    \caption{Observed Purcell enhancement from the Er$^{3+}$-defect: \textbf{a} observed lifetime-transients from the 1541.172 nm transition investigated within all three considered scenarios, where measured data-points (dotted) are fitted with a weighted single-exponential distribution (solid lines) 
    \textbf{b} Overview of all obtained lifetimes, where vertical error-bar indicate the \deleted{fitment} \added{fitting} uncertainty; \textbf{c} Overview of the calculated lifetime-reduction factors $\tau_{\text{thin-film}}/\tau_{\text{on}}$ as well as \deleted{identified} \added{determined} Purcell enhancement factors $\text{F}_{\mathrm{p, Offres}}^{\text{meas.}}$ which takes the off-resonance lifetime into account as well as the determined transition-specific $\text{F}_{\mathrm{p, trans.}}^{\mathrm{meas.}}$ factor\deleted{in relation to the observed associated to the corresponding Er-transition}.}\label{fig3} 
\end{figure}

In addition to \deleted{sufficient} \added{an observed high PL} \deleted{PL} enhancement and \added{remarkably} small spectral diffusion at RT, \added{we subsequently investigated the single Er$^{3+}$ transitions} \deleted{subsequent determined} optical lifetimes \added{which revealed} high lifetime reductions \deleted{which also incline} \added{due to a} significant Purcell enhancement. By exciting at a inhomogeneous peak resonance and recording the subsequent photon-events after excitation-extinction, we observe characteristic single exponential decays for both investigated conditions, `on resonance' and `thin-film'. We detuned the cavity by adjusting the coupling region of the microring and the feed waveguide, referred to here as `off-resonance' (see SI III) and\deleted{and also} record the lifetime transients, illustrated \deleted{with}in Fig. \ref{fig3}\textbf{a}. As shown \deleted{with}in Fig. \ref{fig3}\textbf{b}, individual optical lifetimes of 5.55 $\pm$ 0.08 ms and 5.52 $\pm$ 0.06 ms for the 1538.936 nm (see SI IV) and 1541.172 nm transition, respectively, \added{are measured}, considering \added{the laser} excitation parallel to the \added{main} crystal c-axis in  unpatterned thin-film ($\tau_{\text{thin-film}}$) . Perpendicular \added{to the} c-axis \added{laser} excitation \added{coupled with the detuned resonator} provides optical lifetimes ($\tau_{\text{off-res.}}$) of 0.77 $\pm$ 0.04 ms and 1.18 $\pm$ 0.1 ms, while a on-resonance tuned cavity provides measured lifetimes ($\tau_{\text{on-res.}}$) of 49.95 $\pm$ 2.74 $\mu$s and 38.25 $\pm$ 1.07 $\mu$s, respectively. \deleted{With this,} We determine the lifetime reduction (LR) factors from the total decay rate $\tau_{\text{thin-film}}/\tau_{\text{on-res.}}$ of $\sim$111 and 144 for both Er$^{3+}$ transition lines and Purcell enhancement factors $\text{F}_{\mathrm{p, Offres}}^{\text{meas.}}$ of $\sim$51.9 and 69.8, where a branching ratio (BR) of 0.5~\cite{lyasota2025narrow} and the determined off-resonance lifetimes are considered. More individual Er-transition specific $\text{F}_{\mathrm{p, trans.}}^{\mathrm{meas.}}$-factors can be determined if the BR is utilized to correct the identified LR-factor since the microring selectively enhances only specific transitions. With this assumption, $\text{F}_{\mathrm{p, trans.}}^{\mathrm{meas.}}$ of 222 and 288 can be inferred, also shown in Fig. \ref{fig3}\textbf{c}.

The measured Purcell enhancement places this microring resonator close to previously reported values from optimized Er$^{3+}$-doped \added{photonic crystal (PhC) cavities} in silicon \cite{dibos2022purcell, gritsch2023purcell}; although the Q-factor of the investigated microring is higher, 

\noindent the optical performance is as yet not superseding PhC cavities fabricated in other semiconductors, mostly due to an unoptimized structure for the emission dipoles collection efficiency. However, it outperforms the observed Purcell enhancement of $\sim$50 from a single photon emitting neutral $\text{V}_{\mathrm{Si}}\text{V}_{\mathrm{C}}$ embedded in a 4H-SiC PhC suspended cavity~\cite{crook2020purcell} as well as the most recent demonstration utilizing PL4 $\text{V}_{\mathrm{Si}}\text{V}_{\mathrm{C}}$ \added{in microring resonators in the same material utilized here, i.e. 4H-SiCOI, resulting in } $\text{F}_{\text{p}}$= 5~\cite{bao2025tunable}.

\subsection*{Single photon emission}\label{sec4}
\begin{figure}[H]
    \centering
    \includegraphics[keepaspectratio, width=\textwidth]{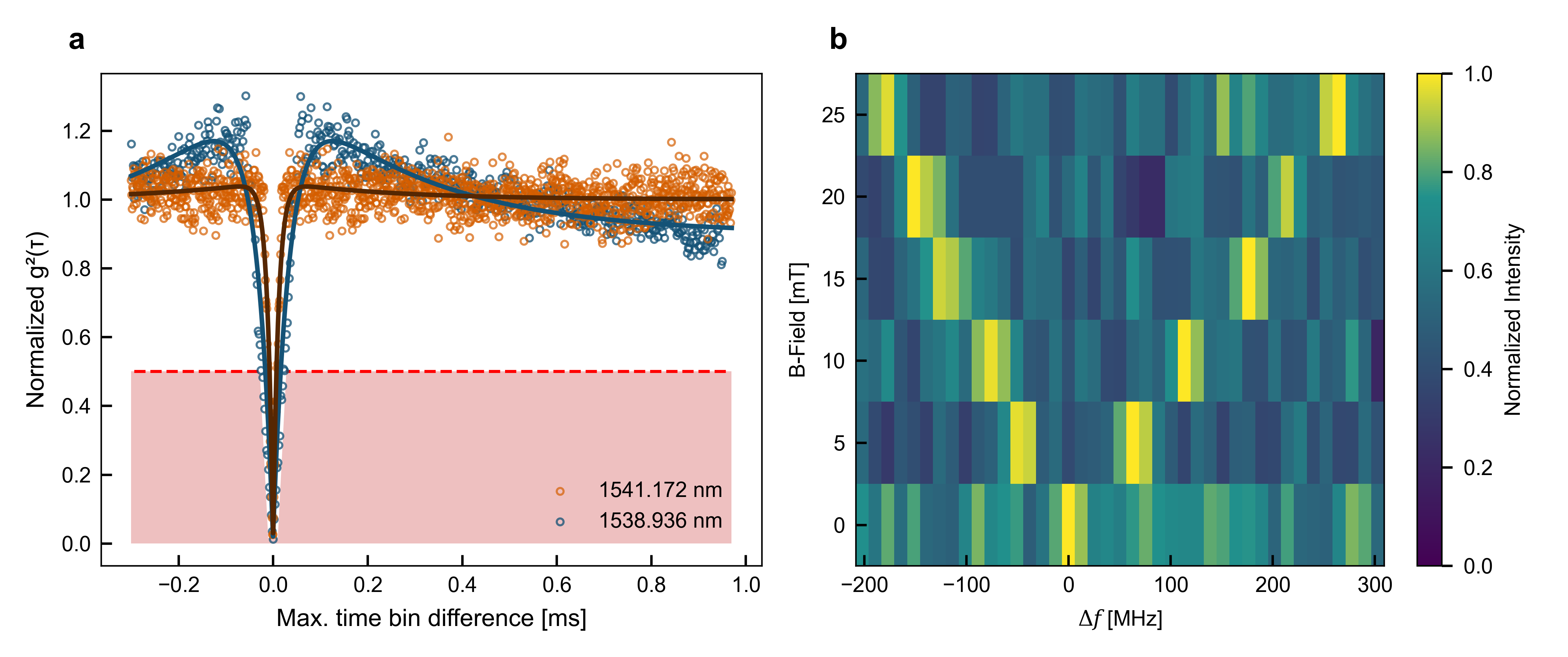}
    \caption{Photon emission autocorrelation measurement and magnetic field dependence  from the implanted Er$^{3+}$-defect within the microring: \textbf{a} maximum time bin difference from the observed 1541.172 nm (orange) and 1538.936 nm (blue) Er$^{3+}$-transition excited with a few hundred nWs on-chip. The red dashed line indicates the critical threshold of 0.5 \added{to identify single photon emission.} Measured datapoints (scattered) are exponentially fitted (solid lines) \textbf{b} Observed magnetic-field dependence from the 1541.172 nm transition where a B-field is applied in 5 mT increments.}  \label{fig4} 
\end{figure}

Furthermore, the observed Purcell enhancement provided by the investigated cavity enables the \deleted{capturing} \added{observation} of single photon emission. For this, rather than implementing the common Hanbury-Brown Twiss setup \cite{brown1956correlation}, we utilized only one detector, registering the photons arrival times since the observed emitter on-resonance lifetimes are several magnitudes higher than the SNSPD dead time of \deleted{has been determined to} less than 30 ns. Via post-processing of the obtained photon arrivals timestamps, we identify distinct antibunching \added{signatures} for both investigated Er$^{3+}$ transition lines, as illustrated in Fig. \ref{fig4}\textbf{a}. Considering the 1541.172 nm transition, a $\text{g}^{(2)}(\tau = 0)$ = 0.04 was identified. In contrast, a specific investigated Er$^{3+}$-transition within unpatterned thin-film 4H-SiCOI sample uniformly implanted with Er$^{3+}$-concentrations an order of magnitude higher ($1.2\times10^{17}$ $\text{Er}/\text{cm}^3$ - See further spectral properties in SI V) than the ones utilized within the microring to isolate single emitters, provides a $\text{g}^{(2)}(\tau = 0)$ = 0.925.

This slightly higher $\text{g}^{(2)}(\tau = 0)$-value exhibited from the microring compared to observed antibunching within a SOI:PhC~\cite{gritsch2023purcell}, can be attributed to a higher mode volume of the \added{microring} cavity. \deleted{However,} \added{Nevertheless,} this \added{result} still shows clear antibunching well below the 0.5 threshold, \deleted{deeming} \added{confirming} the transition as single photon source with negligible photon-emission contribution from other addressed defects which may reside in other parts of the investigated waveguide. 

Since Er$^{3+}$-defects in SiC like in other semiconductors are defined as spin-1/2 system, we seek to investigate these properties and further confirm the observation of individual Er-defects. For this, a magnetic-field (B-field) is applied parallel to the crystals c-axis, which lifts the degeneracy
states from each individual Zeeman-level, leading to theoretically two observable radiative peaks emerging. Localized PLE-measurements were performed while simultaneously increasing the B-field intensity linearly with 5 mT increments after each conducted scan where the observed fluorscence indicates the observation of a Zeeman-split, as shown in Fig. \ref{fig4}\textbf{b}. By correlating each individual peak frequency for each applied B-field step, a $\Delta g$ of 17.59 $\pm$ 0.6 GHz/T can be derived via a linear distribution fit (see SI VI) which points to a spin-preserving Er-transition~\cite{lyasota2025narrow}. Further information could be gained by extracting the complete $g$-tensor which can be identified when the applied B-field is rotated into the remaining two directions. These remain most likely aligned with the basal plane of the crystal assuming a C$_{3v}$ defect-symmetry which could point to a Er$^{3+}$ Si-substitution within the SiC-lattice. Lastly, coherent light was observed when the polarization-dependence is examined (see SI VII).

\end{multicols}
\begin{multicols}{2}

\section*{Discussion}\label{sec5}

In conclusion, we have successfully coupled an ion implanted single Er$^{3+}$-emitter to an optical mode propagating through a microring resonator, which provides multiple sharp spectral resonances with a highest loaded Q-factor of $\sim$$4.26 \times 10^{5}$. Overlapping these resonances with Er$^{3+}$ transition-lines led to significant fluorescence enhancement where spectral diffusion was identified to be as narrow as 54.76 $\pm$ 1.8 MHz.

This spectral diffused linewidth could be further reduced by employing higher purity 4H-SiC (currently based on HPSI substrates), such as from commercial lowest n-doping epilayers \cite{he2024robust} to non-commercial/research grade lowest n-doping epilayers with additional isotopic purification, or employing device engineering such as implementing PiN-junctions near the emitter~\cite{WangOptimizingelectro-optic2025}, removing undesired surrounding charges \cite{anderson2019electrical} and implementing Schottky-diodes in combination with solid immersion lenses~\cite{steidl2025single}.

By investigating the optical lifetime transients from the defect under several conditions, we identify maximum Purcell enhancement of 69.8, assuming a branching ratio of 0.5~\cite{lyasota2025narrow}, which 
\noindent provides an excellent enhancement to observe single photon emission. The here reported Purcell enhancement is almost an order of magnitude higher than the latest values observed within 4H-SiCOI microring resonators, with the highest $\text{F}_{\text{p}}$ of 5 reported for PL4 divacancies~\cite{bao2025tunable}. The single-photon emission remains well below the 0.5-threshold at 0.04. However, optimized PhC cavities in silicon have provided purer antibunching of Er$^{3+}$-defects at $\text{g}^{(2)}(\tau = 0)$ of 0.017~\cite{gritsch2023purcell} and 0.019~\cite{gritsch2025optical} with a full comparison from the most recent demonstrations presented in Table \ref{tab:Result_Comp}. We attribute this higher $\text{g}^{(2)}$-value to the higher mode volume from the microring resonator compared to PhC cavities which may be further reduced by spatially separated implantation via for example an implantation mask~\cite{babin2022fabrication}, by reducing the mode volume of the investigated cavity or further reducing the present dark-counts from the SNSPD.

All in all, we utilize a scalable material platform and approach for embedding the Er$^{3+}$-defect into a micro-cavity, which allows the \deleted{capture} \added{isolation and detection} of single photons. \deleted{That} \added{This result} shows a scalable way for quantum technology application of Er$^{3+}$-defects embedded in 4H-SiCOI at RT. Further investigations are required to prove indistinguishable photons emission.
A previous report shows single indistinguishable photons from Er$^{3+}$ in a $\text{CaWO}_{4}$:PhC with an Er$^{3+}$ concentration of $5\times10^{14}$ $\text{Er}/\text{cm}^3$, which is 1.5 times lower than the one \deleted{proposed} \added{presented in this work} \deleted{here} for single photon emission \cite{ourari2023indistinguishable}. \deleted{Also,} \added{Finally, optically detected magnetic resonance} single shot read-out could be studied \added{in this platform using either a PhC or a microring resonator at cryogenic temperatures to}\deleted{for} further \added{advance the }quantum technology applications of Er$^{3+}$. 
These results pave the way for Er$^{3+}$-defects within 4H-SiCOI as a promising candidate for scalable spin-photon interfaces operating at the single photon level at higher temperature and provide opportunities to further enhance other color centres in 4H-SiCOI photonics.

\end{multicols}
\begin{table}[!t]
  \centering
  \begin{tabular}{l c c c c c c c}
    \hline
    Host & Defect & Cav.-type & Q-fact. & Spect. broaden. &$\text{F}_{\text{p}}$ & $\text{g}^{(2)}(\tau = 0)$ & Ref. \\
    \hline
    \noalign{\vskip 2pt}
    SOI & Er$^{3+}$ & PhC. & $1.08\times 10^{5}$ & 79 MHz & 78.4 & 0.017 &~\cite{gritsch2023purcell}\\
    SOI & Er$^{3+}$ & PhC. & $8.2\times 10^{4}$ & 47 MHz & 177 & 0.019 &~\cite{gritsch2025optical}\\
    YSO & Er$^{3+}$ & Fab.-Per. & $9 \times 10^{6}$ & - & 59 & - &~\cite{merkel2020coherent}\\
    $\text{TiO}_{2}$ & Er$^{3+}$ & PhC. & $5.3 \times 10^{4}$ & 52 GHz & $<$200 & - &~\cite{dibos2022purcell}\\
    $\text{CaWO}_{4}$ & Er$^{3+}$ & PhC. & $1.9 \times 10^{5}$ & $\sim$150 kHz & 850 & 0.018 &~\cite{ourari2023indistinguishable}\\
    4H-SiC & $\text{V}_{\mathrm{Si}}\text{V}_{\mathrm{C}}$ (neutr.)  & PhC. & $5.1 \times 10^{3}$ & 3.98 GHz & 50 & 0.096 & ~\cite{crook2020purcell} \\
    4H-SiC & $\text{V}_{\text{Si}}^{-}$ & PhC. & $5.3 \times 10^{3}$  & - & 89 & - & ~\cite{bracher2017selective} \\
    4H-SiCOI & $\text{V}_{\text{Si}}^{-}$ & PhC. & $1.49 \times 10^{4}$ & - & $>$9 & 0.08 & ~\cite{lukin20204h}\\
    4H-SiCOI &  $\text{V}_{\mathrm{Si}}\text{V}_{\mathrm{C}}$ (PL4) & Microring & $1.2 \times 10^{3}$ & - & 5 & - & ~\cite{bao2025tunable}\\
    4H-SiCOI & Er$^{3+}$ & Microring & $4.26 \times 10^{5}$ & $\sim$54 MHz & 69.8 & 0.04 & (this work at RT)\\
    \hline
  \end{tabular}
  
  \caption{Summary of most recent studies on Er$^{3+}$ incorporated in photonic or optical cavities in various materials and of the integration of color centres in SiC and 4H-SiCOI photonics, selected on the basis of Purcell enhancement, spectral and single photon emission purity as they compare with this work, the only one performed at RT.} 
  \label{tab:Result_Comp}
\end{table}
\begin{multicols}{2}
\section*{Methods}\label{sec7}
\subsection*{Sample preparation}\label{sec8}

The microring resonator is fabricated \added{on 4H-SiCOI} in a similar manner as reported in~\cite{cai2022octave}, utilizing wafer scale  High Purity Semi-insulating (HPSI) 4H-SiC from CREE and processed into 4H-SiCOI using a silicon handle for wafer bonding and subsequent thinning to 630 nm using chemical-mechanical polishing and reactive ions etching~\cite{cai2022octave}.
By performing a parameter-sweep with subsequent characterization, we identify acceptable interaction between resonator-mode and Er-defect within a waveguide widths (WW) of 1.5 $\mu$m. Microring waveguide widths (RW)  modifications from 2.2 $\mu$m \noindent in a on-resonance state to 2 $\mu$m detunes the cavity modes sufficiently to an off-resonance state.
The investigated microrings with a peak concentration of $1.2\times10^{16}$ $\text{Er}/\text{cm}^3$ were prepared by a two-step ion implantation approach as reported in Refs.~\cite{bader2025photo, lyasota2025narrow} where the peak ion distribution is located at a depth region between 250-300 nm. An ion fluence sweep was performed to study the impact of different ion fluences onto the spectral response. Individual ion fluences of $5\times10^{10}$, $1\times10^{12}$ and $1\times10^{14}$ $\text{Er}/\text{cm}^2$ were utilized with ion implantation energies of 350 keV under ion channeling conditions. 
Each individual sample was subject to a 30 minute anneal at 1000\textdegree C in a Argon-atmosphere after fabrication and implantation. We note that \deleted{this annealing process} \added{the maximum annealing temperature} is limited due to the $\text{SiO}_{2}$/Si interface which delaminate at higher temperatures.

\subsection*{Optical analysis}\label{sec8}

Each spectral response from the microring is captured via a V-grooved fiber array \deleted{which is} aligned with the uniform grating couplers via Luminos I6000 nanopositioners. A instantaneously tunable 1 mW laser-input (HP81640A) is then coupled into the device and the transmitted output-power measured via an optical power-meter (HP 81532A).
The measured power is converted to transmission in dB via postprocessing. Subsequent single-lorentzian fits are applied to identify the individual FWHMs from each resonance.

PLE-scans are carried out with a continuous-wave excitation laser (\deleted{provided by a}HP81640A or ID Photonics Cobrite DX1)  \deleted{applied instantaneously} tuned from 1530 nm to 1555 nm (195.942 THz to 192.792 THz) with various increments (12.5 MHz or 62.5 MHz). This excitation is subsequently modulated with acoustic-optic modulator (AOM) which provides 45 dB extinction ratios in contrast between `on' and `off' in a free-space configuration. After the laser pulse excites the Er$^{3+}$ emitter, an SNSPD (Single Quantum Eos 810)  detects the incoming emitted photons. The SNSPD bias-current is also modulated during excitation-events to prevent latching-occurrence. The detected photons are then subsequently timestamped in conjunction with the falling edge of the excitation event by a ID Quantique ID801 timetagger with a time resolution of 81 ps. Via post-processing, we neglect all events which may have occurred before the falling edge and extract all counts within a specific timeframe which depends on the application after excitation-extinction. 
The described mechanism is then applied in two ways: (1) Via the previously mentioned V-grooved fiber-array to address the Er$^{3+}$-emitter within the microring perpendicular to the crystal c-axis. This technique is illustrated in Fig. \ref{fig2}\textbf{a}; (2) in a free-space confocal configuration in parallel to the crystals c-axis via a LCPlan 100x 0.85NA Olympus objective in combination with a 10:90 beam splitter (Thorlabs BSN12 10:90 R:T) .  

After the PLE-scans, the inhomogeneous peaks can be identified via single Gaussian fits. We identified these peaks, center the tunable excitation on these and then drive the transition. The obtained data is then processed, normalized to the trigger event and binned where single exponential fits then reveal the exhibited lifetime. 
By correlating the timetagged arrival-time from the captured photons against each other, we identify clear antibunching dips. Considering the examined Er-defects within the microring, this overall correlation time is set to 1 ms which covers the observed reduced on-resonance lifetime well with a subsequently chosen bin-size of 1 $\mu$s.

Furthermore, the measured Purcell enhancement can be determined via the obtained lifetimes as \cite{bao2025tunable}, 
\begin{equation}
\text{F}_{\mathrm{p, Offres}}^{\text{meas.}}
= \frac{\tau_{\text{thin-film}}}{\tau_{\text{off-res.}}}
\left( \frac{\tau_{\text{off-res.}}}{\tau_{\text{on-res.}}} - 1 \right)\, \times DWF,
\end{equation}

where the Debye-Waller Factor (DWF) can be replaced by the branching ratio (BR) from the observed defect, based on previous PLE-investigations~\cite{gritsch2022narrow}. Lifetime reductions are determined as the ratio between $\tau_{\text{thin-film}}$ and $\tau_{\text{on-res.}}$ with the corrected measured $\text{F}_\text{P, trans.}$-factor determined with Eq. (2).

\begin{equation}
\text{F}_{\mathrm{p, trans.}}^{\mathrm{meas.}}
= \frac{\tau_{\mathrm{thin-film}}}{\text{BR}\times \tau_{\mathrm{on-res.}}}
\end{equation}


Lastly, the observed antibunching-traces are fitted via Eq. (3),
\begin{equation}
g_2(\tau) = 1 - (1 + a) \, e^{- \frac{|\tau|}{\tau_1}} + a \, e^{- \frac{|\tau|}{\tau_2}}
\end{equation}

\backmatter

\section*{Acknowledgements}

We acknowledge the NCRIS Heavy Ion Accelerator platform (HIA) for access and support to the ion implantation equipment at the Australian National University.

This work was performed in part at the RMIT Micro Nano Research Facility (MNRF) in the Victorian Node of the Australian National 
Fabrication Facility (ANFF) and the RMIT Microscopy and Microanalysis Facility (RMMF).

S.-I.S. acknowledges the financial support provided by the JST FOREST Program (Grant No. JPMJFR203G), the JSPS KAKENHI (Grant No. JP22H03880), and the QST grants-in-Aid for Exploratory Research. 

J.McC. acknowledges the Australian Government Australian Research Council under the Centre of Excellence scheme (No: CE170100012). 

A.L. and S.R. are supported by the ARC Centre of Excellence for Quantum Computation and Communication Technology (Grant CE170100012) and the Discovery Project (Grant DP210101784).

Q.L. is supported by the National Science Foundation of the United States of America under Grant No. 2240420.

J.B. acknowledges the help of Hassan J. Latief and Nikita Komarov who contributed to the experimental analysis.

All authors acknowledge the work from Jingwei Li, who contributed to the fabrication of the investigated samples.

\section*{Authors contribution}

Conceptualization: S.C., J.McC., Q.L., J.B., A.L., S.-I S.; Data curation: J.B.; Formal analysis: J.B., S.C.; Investigation: J.B., A.L., J.McC., S.-I S.; Methodology: J.B., A.L., S.C., S.-I S., S.Q.L., R. W., J.McC., Q.L.; Resources: S.C., S.R., S.Q.L., J.McC., R.W., Q.L., S.-I S., D.B.; Supervision: S.C., B.C.J.; Visualization: J.B.; Writing - original draft: J.B., S.C.; Writing - review \& editing: J.B., A.L., B.C.J., Q.L., S.-I S., D.B., S.Q.L., S.R. All authors have read and agreed to the published version of the manuscript.

\section*{Competing Interests}

The authors have no competing interests

\section*{Data availability}

The presented data is available upon reasonable request from the corresponding author.
\end{multicols}


\begin{multicols}{2}
\bibliography{sn-bibliography}
\end{multicols}
\newpage

\renewcommand{\thefigure}{S\arabic{figure}}
\renewcommand{\thetable}{S\arabic{table}}
\raggedbottom

\section*{Supplementary Information: Room temperature Purcell enhanced single erbium ions in silicon-carbide-on-insulator microring resonators}

\fnm{Joshua} \sur{Bader}
, \fnm{Shin-ichiro} \sur{Sato}, \fnm{Jeffrey C.} \sur{McCallum},
\fnm{Ruixuan} \sur{Wang}, \fnm{Shao Qi} \sur{Lim}, \fnm{Alexey} \sur{Lyasota},
\fnm{David} \sur{Broadway}, \fnm{Brett C.} \sur{Johnson}, \fnm{Sven} \sur{Rogge},
\fnm{Qing} \sur{Li}, \fnm{Stefania} \sur{Castelletto}



\affil[1]{\orgdiv{School of Engineering}, \orgname{RMIT University}, \orgaddress{\city{Melbourne}, \postcode{3000}, \state{VIC}, \country{Australia}}}
\affil[2]{\orgdiv{National Institutes for Quantum Science and Technology}, \orgaddress{\city{Takasaki, Gunma}, \postcode{370-1292}, \country{Japan}}}
\affil[3]{\orgdiv{Centre for Quantum Computation and Communication Technology, School of Physics}, \orgname{The University of Melbourne}, \orgaddress{\city{Melbourne}, \postcode{3010}, \state{VIC}, \country{Australia}}}
\affil[4]{\orgdiv{Electrical and Computer Engineering}, \orgname{Carnegie Mellon University}, \orgaddress{\city{Pittsburgh}, \postcode{15213}, \state{PA}, \country{USA}}}
\affil[5]{\orgdiv{School of Science}, \orgname{RMIT University}, \orgaddress{\city{Melbourne}, \postcode{3001}, \state{VIC}, \country{Australia}}}
\affil[6]{\orgdiv{Centre of Excellence for Quantum Computation and Communication
Technology, School of Physics}, \orgname{University of New South Wales}, \orgaddress{\city{Sydney}, \postcode{2052}, \state{NSW}, \country{Australia}}}



\maketitle

\tableofcontents

\section*{Introduction}
In this Supplementary Information, we show additional data supporting our conclusions, presented in the main manuscript. In Section I, we provide further insights into the conducted microring spectral response and provide Lorentzian fits which are utilized for FWHM extractions. Section II shows a study where increasing Er ion fluence and their subsequent impact onto the spectral response from a $\sim$70$\mu$m diameter microring are investigated.
A detuned cavity, further referenced as `off-resonance' is then introduced in section III with insights into the spectral response from the microring and subsequently obtained PLE-spectra. 
A second successfully coupled Er-transition is then introduced in section IV with presented results into lifetime-dependent PLE-spectra as well as optical lifetime transients.
In section V, the identified peak-frequency shifts under an increasing magnetic field are presented where a linear distribution is fitted. That allows us to derive the $\Delta$g entry. 
In section VI, the impact of polarization dependent excitation is shown. 
\addcontentsline{toc}{section}{Section I. PIC spectral response - Lorentzian fits}
\section*{Section I. PIC spectral response - Lorentzian fits}
\begin{figure}[H]
    \centering
    \includegraphics[width=\textwidth]{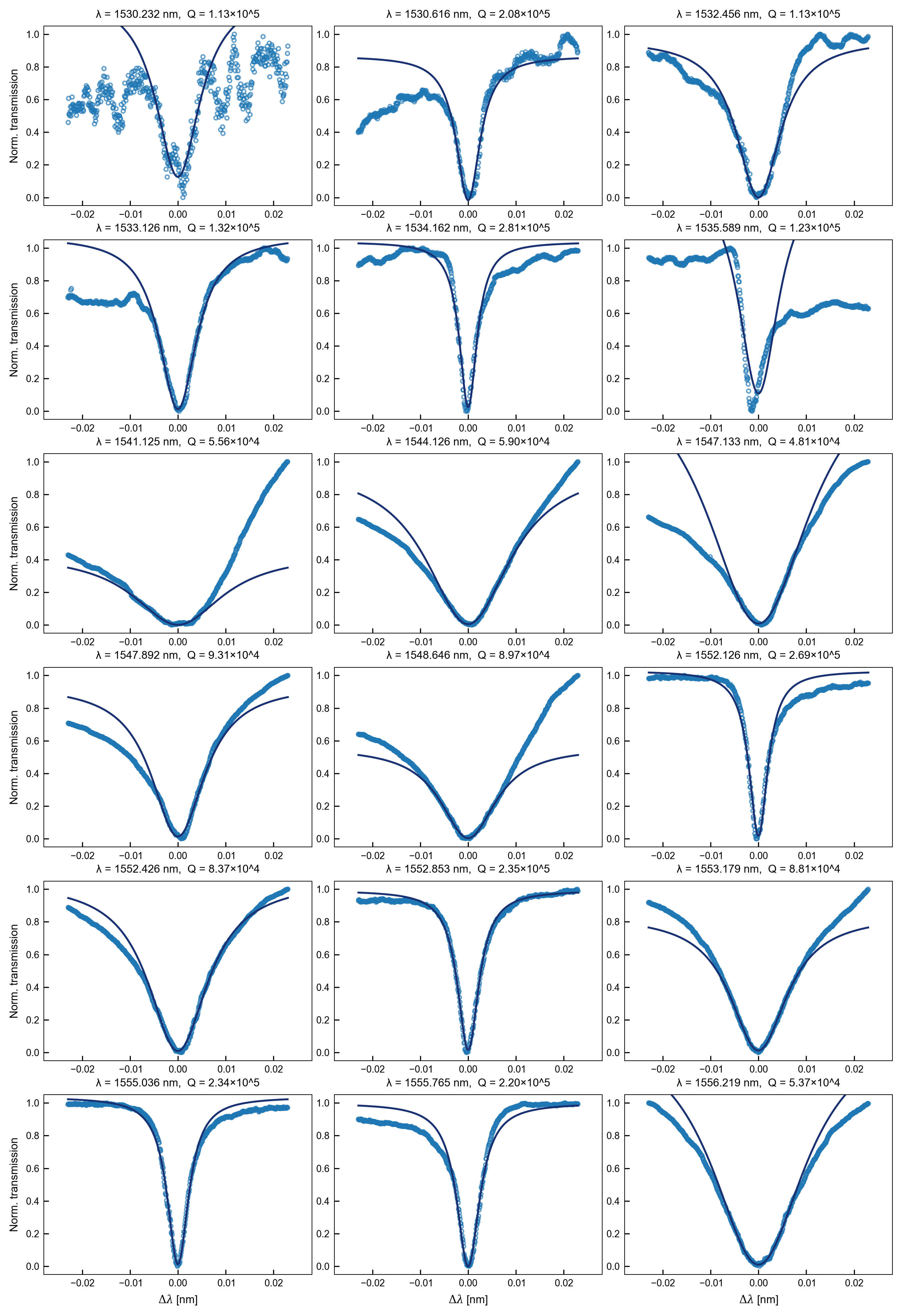}
    \caption{Fitted identified resonances from the microring.}
    \label{fig:Comp_ImplDose}
\end{figure}

\addcontentsline{toc}{section}{Section II. Impact of the Er ion fluence on the spectral response}
\section*{Section II. Impact of the Er ion fluence on the spectral response}

Fig. \ref{fig:Comp_ImplDose} shows the investigated implantation dose sweep with shallow Er-emitters. As shown in Table~\ref{tab:implant_Q}, Lorentzian fits reveal decreasing Q-Factor for higher Er ion fluences, indicating that the present Er-dopants can deteriorate the propagating mode in the microring.

\begin{figure}[H]
    \centering
    \includegraphics[width=0.65\textwidth]{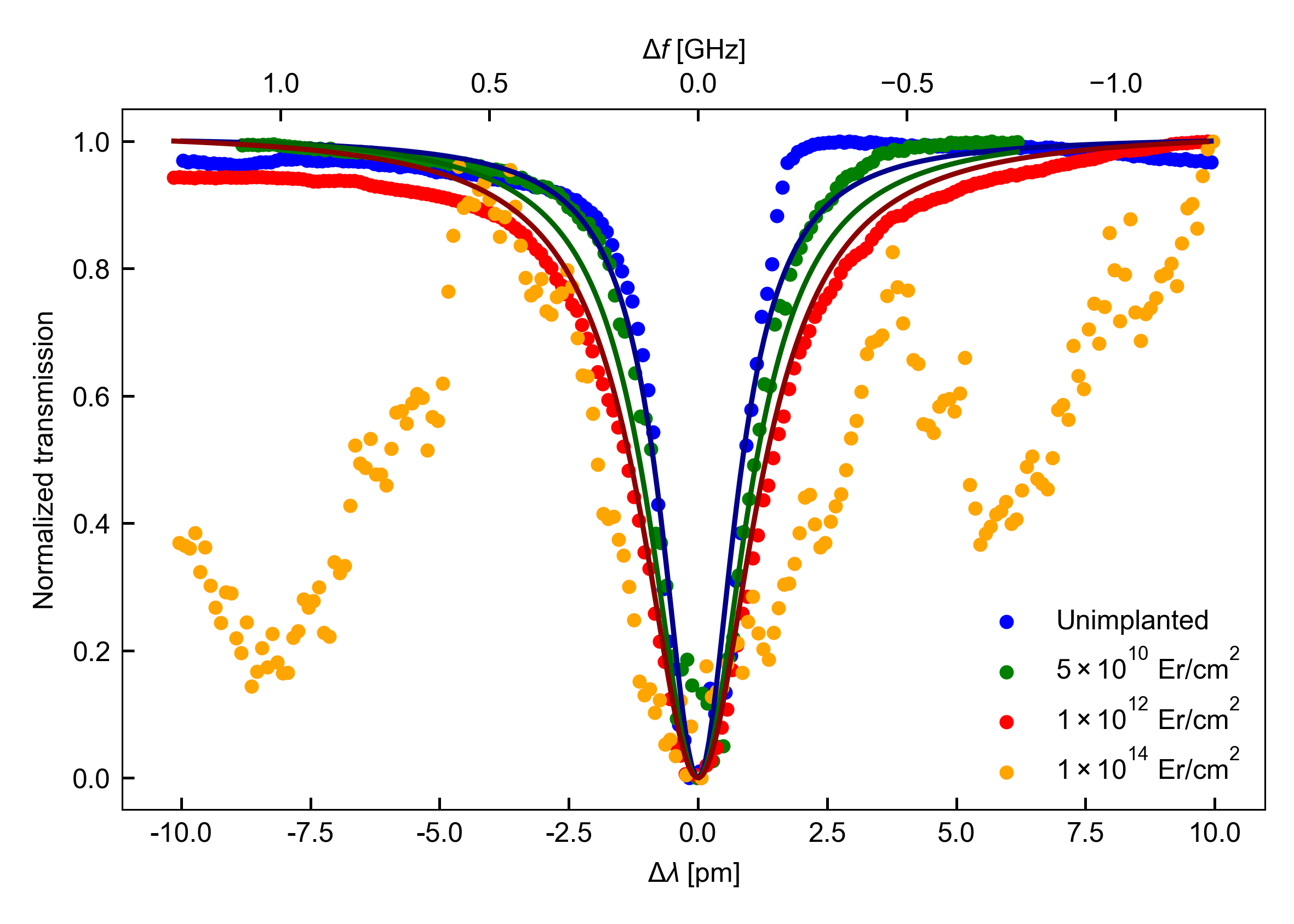}
    \caption{Comparison between a 70 $\mu$m microring implanted with different Er implantation doses.}
    \label{fig:Comp_ImplDose}
\end{figure}

\begin{table}[ht]
\centering
\caption{Overview of specific resonances from different Er-doses within 70 $\mu$m microring resonators}
\label{tab:implant_Q}
\begin{tabular}{lccc}
\hline
\textbf{Sample} &
\textbf{FWHM [MHz]} &
\textbf{$Q$-factor} \\
\hline
Not implanted &
215.6 $\pm$ 9.6 &
894\,894.2 \\

$5\times10^{15}$ Er/cm$^{3}$ &
286.3 $\pm$ 19.7 &
674\,225.9 \\

$1\times10^{17}$ Er/cm$^{3}$ &
331.7 $\pm$ 4.3&
581\,840.6 \\

$1\times10^{19}$ Er/cm$^{3}$ &
- &
- \\
\hline
\end{tabular}
\end{table}

\addcontentsline{toc}{section}{Section III. Off-resonance cavity investigation}
\section*{Section III. Off-resonance cavity investigation}
\begin{figure}[H]
    \centering
    \includegraphics[width=0.85\textwidth]{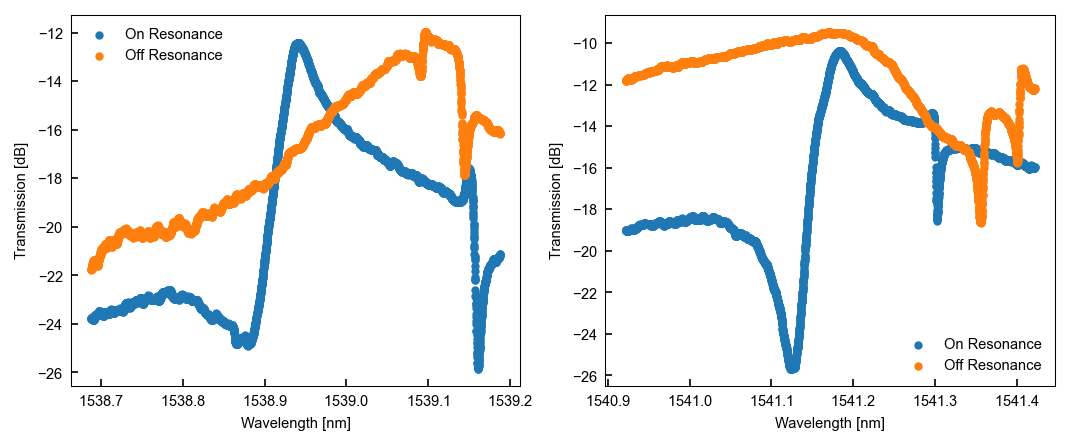}
    \caption{Microring investigation overlapped from the 'on-resonance' and 'off-resonance' cavity}
    \label{fig:MR_SR_Overlap}
\end{figure}

\begin{figure}[H]
    \centering
    \includegraphics[width=0.85\textwidth]{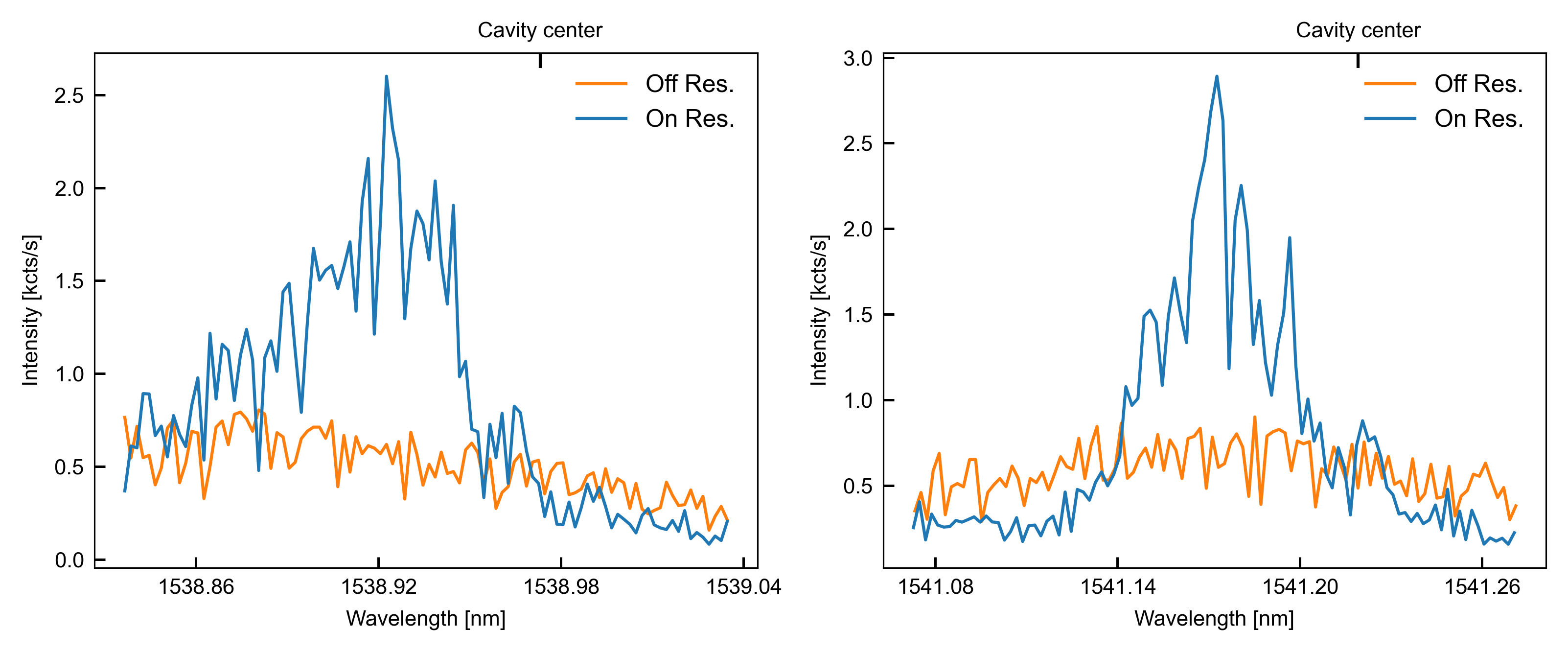}
    \caption{PLE-comparison for on- and off-resonance for 1538.936nm and 1541.172nm transitions.}
    \label{fig:MR_PLE_Overlap}
\end{figure}

In Figure \ref{fig:MR_SR_Overlap}, we investigate the cavity in a detuned state (orange dots)and compare it with the measurements from the on-resonance cavity (blue dots). Our measurements indicate that the optical resonances from the microring are shifted and outside the investigated area from the PLE-scan which is beneficial to examine the defect-properties in a  waveguide without active mode-overlap. 
In Figure \ref{fig:MR_PLE_Overlap}, the obtained PLE-spectra from both cavities are overlapped. While the detuned cavity (orange line) shows no significant enhancement, the on-resonance spectras (blue line) shows strong PL-enhancement within close proximity of the cavity center. All spectras take photon-events 1ms after excitation-extinction into account.

\addcontentsline{toc}{section}{Section IV. 1538.936nm on-resonance spectral properties}
\section*{Section IV. 1538.936nm on-resonance spectral properties}

\begin{figure}[H]
    \centering
    \includegraphics[width=0.5\textwidth]{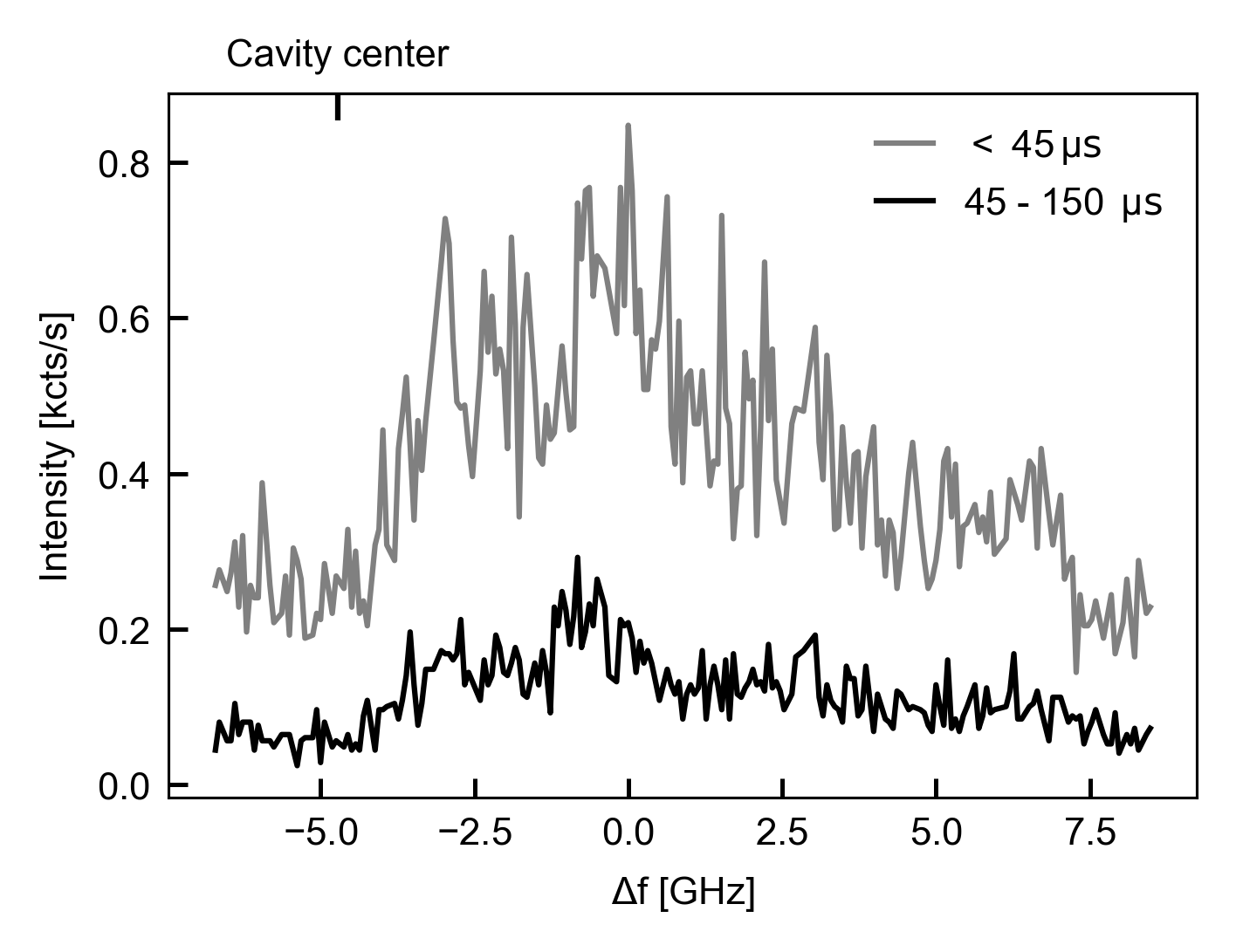}
    \caption{Lifetime dependent PLE from the observed 1538.936nm Er-transition which is here centered at 0 GHz.}
    \label{fig:LT_PLE_1538nm}
\end{figure}

\begin{figure}[H]
    \centering
    \includegraphics[width=0.5\textwidth]{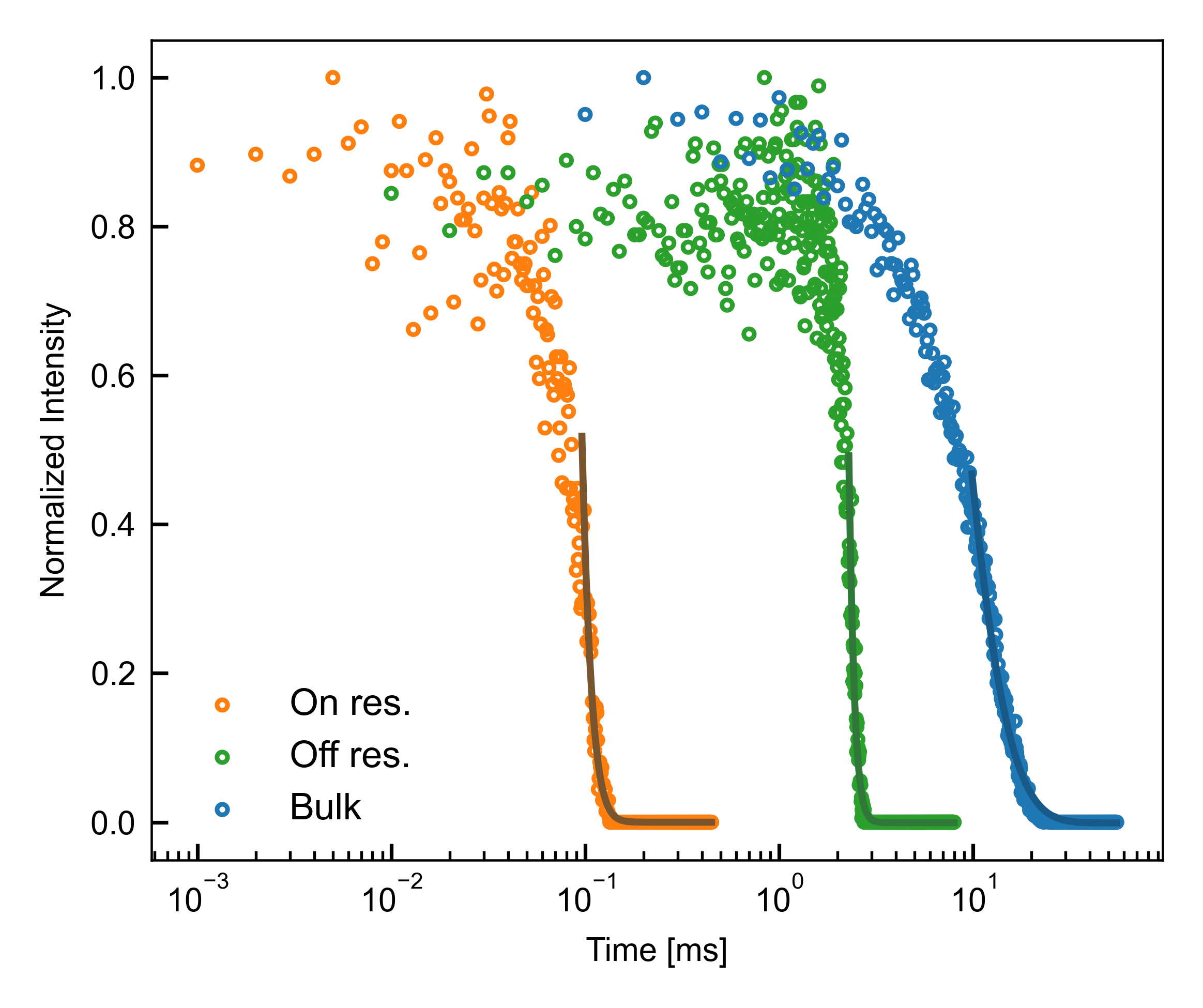}
    \caption{Lifetime transients from the 1538.936 nm Er-transition where measured data (scattered) is single exponentially fitted (solid line).}
    \label{fig:LT_transients_1538nm}
\end{figure}

Figure \ref{fig:LT_PLE_1538nm}, shows the lifetime-dependency from the 1538.936nm Er-transition. Since the optical lifetime is determined to $\sim$49.95 $\mu$s, we expect additional photon-events if the timeframe of 45-150$\mu$s (black line) after excitation-extinction is considered. However, the majority of the emitted photons are still captured within the first 45$\mu$s (grey line).
A subsequent optical lifetime investigation was then conducted, as shown in Figure \ref{fig:LT_transients_1538nm}. Single-exponential fits reveal optical lifetimes of 49.95 $\pm$ 2.74 $\mu$s, 0.77 $\pm$ 0.04 ms and 5.55 $\pm$ 0.08 ms for an on-resonance cavity, off-resonance cavity and bulk 4H-SiCOI, respectively. 

\addcontentsline{toc}{section}{Section V. PLE-reference to $1.2\times10^{17}$ $\text{Er}/\text{cm}^3$ bulk SiCOI}
\section*{Section V. PLE-reference to $1.2\times10^{17}$ $\text{Er}/\text{cm}^3$ bulk SiCOI}

\begin{figure}[H]
    \centering
    \includegraphics[width=0.9\textwidth]{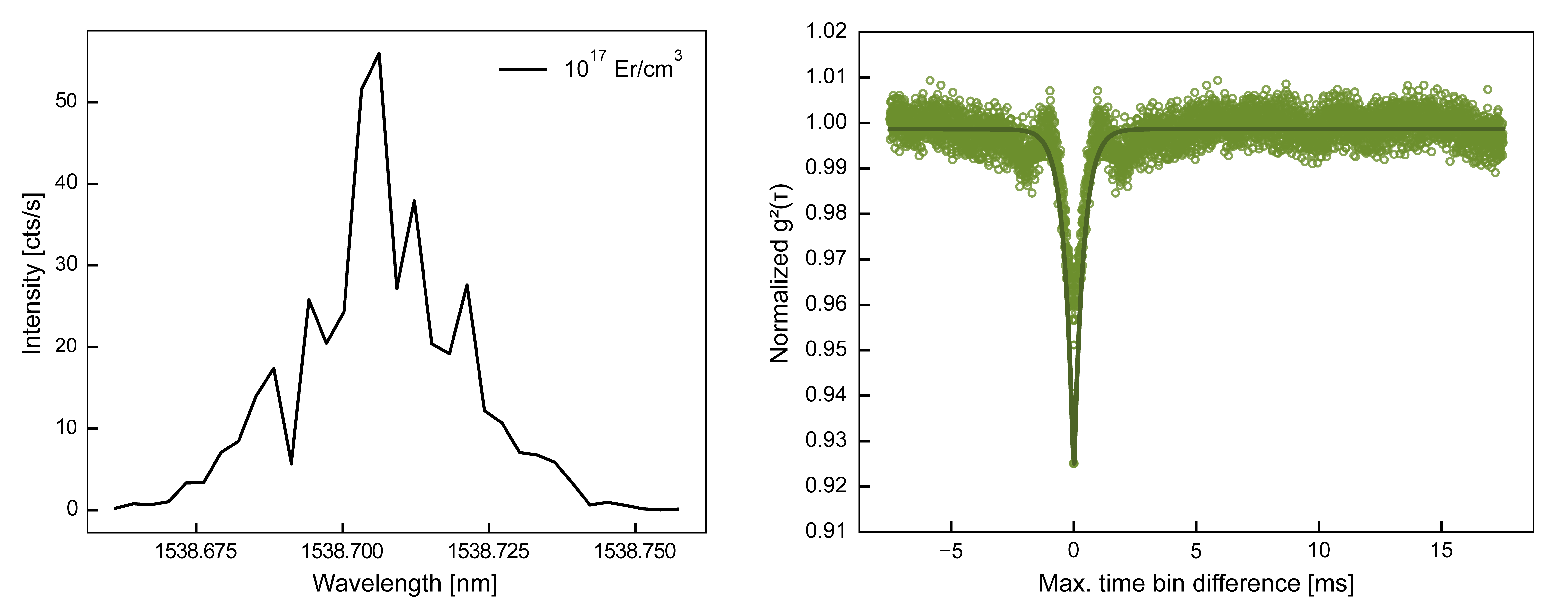}
    \caption{PLE scan conducted with 500MHz increments on a $1.2\times10^{17}$ $\text{Er}/\text{cm}^3$ implanted 4H-SiCOI sample where a FWHM of 33.12 $\pm$ 3.82 GHz is identified.}
    \label{fig:Er12_PLE}
\end{figure}

The optical properties from a bulk $1.2\times10^{17}$ $\text{Er}/\text{cm}^3$ implanted 4H-SiCOI sample are presented here. We find a distinct peak within a PLE-scan at 1538.71 nm  which is shown in Fig. \ref{fig:Er12_PLE}.    

\addcontentsline{toc}{section}{Section VI. B-Field dependence - $\Delta$g fit}
\section*{Section V. B-Field dependence - $\Delta$g fit}
\begin{figure}[H]
    \centering
    \includegraphics[width=0.4\textwidth]{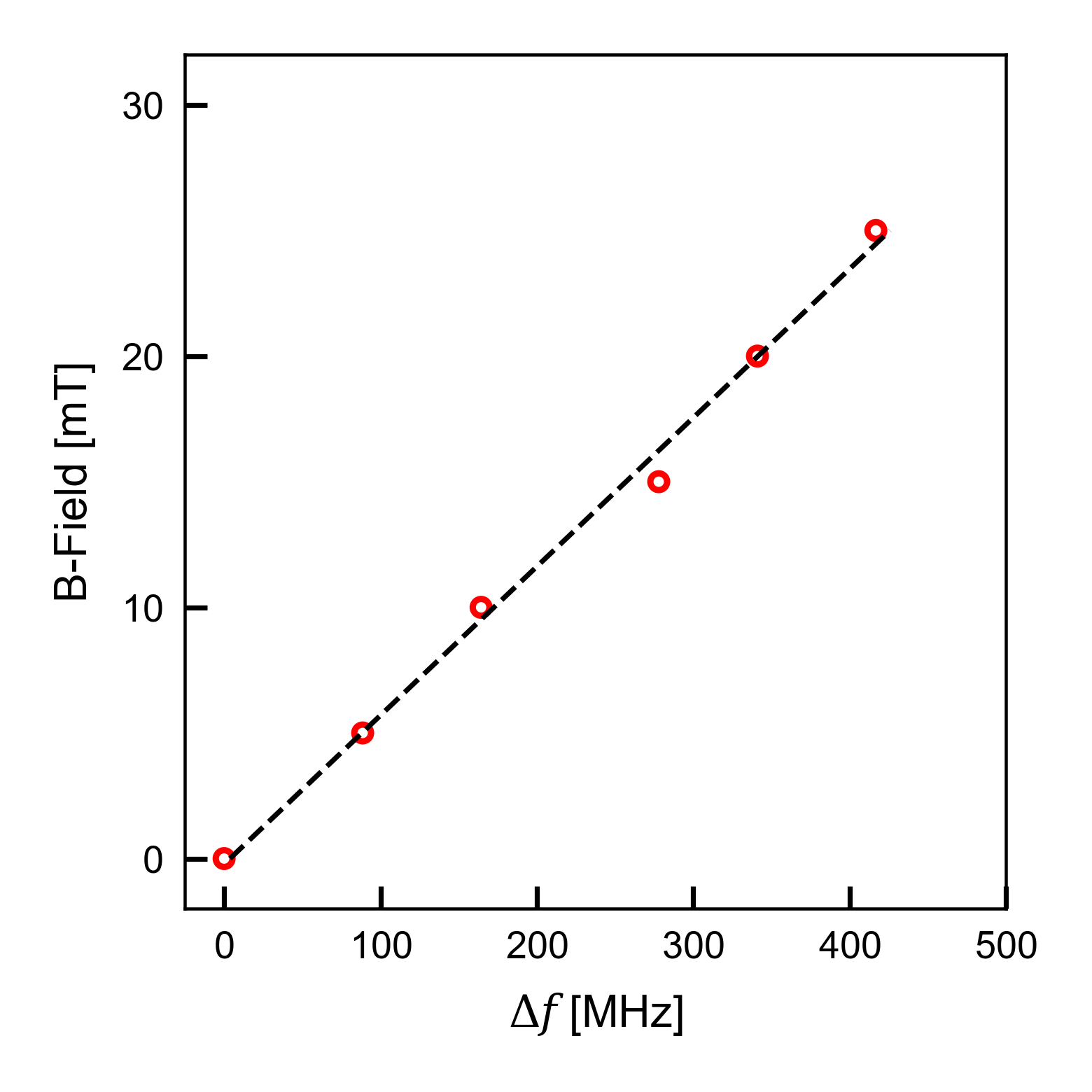}
    \caption{Extracted peak-shift summed up for both  measured emerging arms (red dots) where a linear distribution is fitted (black dashed line)}
    \label{fig:Zeeman_Fit}
\end{figure}

The magnetic-field dependency can be expressed via the determined $\Delta$g factor which can be derived from a linear fit of the obtained frequency shift as a function of increased magnetic field.
As shown in Fig. \ref{fig:Zeeman_Fit}, we identify a clear linear relationship from the measured data (red points) where $\Delta$g is obtained from the slope of the linear fit.

\addcontentsline{toc}{section}{Section VII. Examined polarization dependence}
\section*{Section VI. Examined polarization dependence}

\begin{figure}[H]
    \centering
    \includegraphics[width=0.5\textwidth]{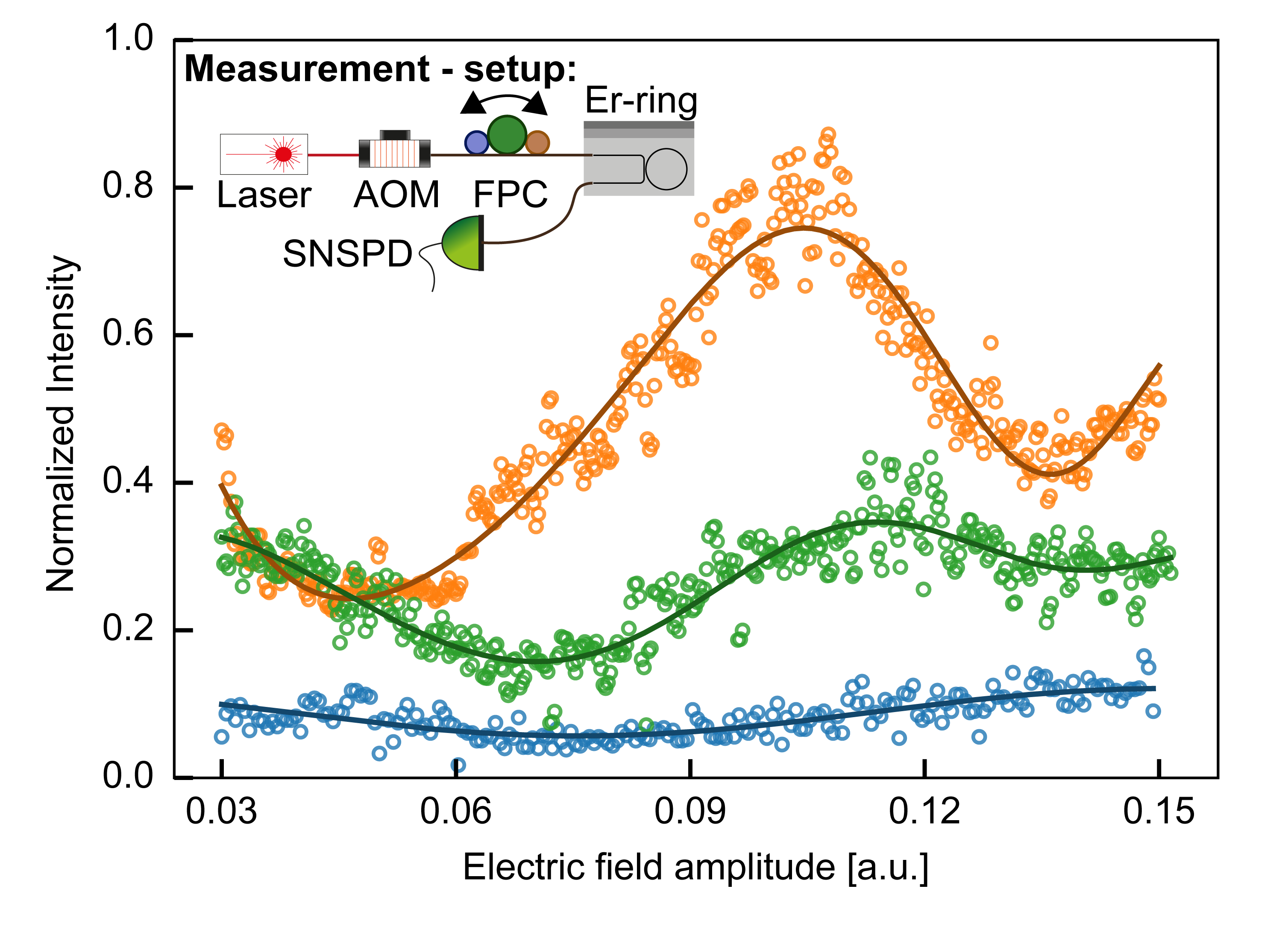}
    \caption{Examined photon-emission probability by tuning the excitation-polarization (dots) via a fiber-coupled polarization controller (FPC) which exhibit fitted Rabi oscillations (solid lines). The inset illustrates the utilized measurement-setup for this investigation.}
    \label{fig:Polarization_Dep}
\end{figure}

Since the excitation of the Er-emitter and 
the coupling of the microring resonator mode is highly dependent on the polarization, we can expect selective Rabi oscillations. As shown within Fig. \ref{fig:Polarization_Dep}, a polarization match reduction (green and blue data) leads to lengthened oscillation-periods while sufficient coupling (orange data) results in coherent quadratic Rabi-oscillations.



\end{document}